\documentstyle[preprint,aps]{revtex}

\tightenlines

\newcommand{\ket}[1]{|{#1}\rangle}
\newcommand{\bra}[1]{\langle{#1}|}

\begin{document}

\title{The Mach-Zehnder and the Teleporter}
\author{T.C.Ralph}
\address{Department of Physics, Faculty of
Science, \\ The Australian National University, \\ ACT 0200 Australia \\ 
 E-mail: Timothy.Ralph@anu.edu.au}
\maketitle

\begin{center}
\scriptsize (June 1999)
\end{center}

\begin{abstract}

We suggest a self-testing teleportation configuration for photon 
q-bits based on a Mach-Zehnder interferometer. That is, 
Bob can tell how well the input state 
has been teleported without knowing what that input state was. One could 
imagine building a ``locked" teleporter based on this configuration. The 
analysis is performed for continuous variable teleportation but the 
arrangement could 
equally be applied to discrete manipulations.

\end{abstract}

\vspace{10 mm}

One problem with teleportation experiments as they are currently performed 
\cite{bou97,bos98,fur98} is that Victor (the verifier) 
must examine the teleported state to 
determine if the machine is working. Victor prepares the original input 
state and is the only person who knows its identity. Because of the 
imperfect nature of experiments even Victor must be careful not to be 
tricked in deciding if some level of teleportation has occurred. 
In principle one might 
imagine checking the teleporter is working once, then leaving it to go, but 
in practice machines drift. Thus it would be useful if a constant, 
straightforward assessment of the teleportation could be carried out without 
prior knowledge of the input.

Consider first the set-up shown schematically in Fig.1(a). Basically we 
place a teleporter in one arm of a Mach-Zehnder interferometer, inject 
a single photon state, in an arbitrary polarization superposition 
state into one port, then use the interference visibility at the 
output ports to characterize the efficacy of teleportation. The beauty 
of such a set-up is the visibility does not depend on the input state, 
so we can assess how well the teleporter is working without knowing 
what is going into it. Let us see how this works.

The input for one port of the interferometer is in the arbitrary polarization 
superposition state
\begin{equation}
\ket{\phi}_{a}={{1}\over{\sqrt{2}}}(x\ket{1,0}+y\ket{0,1})
\end{equation}
where $\ket{n_{h},n_{v}}\equiv \ket{n_{h}}_{h}\otimes 
\ket{n_{v}}_{v}$, $n_{h}$ and $n_{v}$ are the photon number in the 
horizontal and vertical polarizations respectively, and 
$|x|^2+|y|^2=1$. The input of the other port is in the vacuum 
state $\ket{\phi}_{b}=\ket{0,0}$. 
The Heisenberg picture operators for the four input 
modes (two spatial times two polarization) are $a_{h}$ and $a_{v}$ 
(superposition), and $b_{h}$ and $b_{v}$ (vacuum). We propagate these 
operators through the Mach-Zehnder (including the teleporter). After the first 
beamsplitter we can write
\begin{eqnarray}
c_{h,v} & = & {{1}\over{\sqrt{2}}}(a_{h,v}+b_{h,v})\nonumber\\
d_{h,v} & = & {{1}\over{\sqrt{2}}}(a_{h,v}-b_{h,v})
\end{eqnarray}
One of the beams ($c$) is then teleported using the continuous variable 
method discussed in Reference \cite{pol99,ral99}. The individual 
polarization modes of $c$ are separated using a polarizing beamsplitter. 
Each mode is then mixed on a 50:50 beamsplitter with a correspondingly 
polarized member of an entangled pair of beams. The entangled pairs 
may come from two separate 2-mode squeezers or alternatively a single 
polarization/number entangler could be used \cite{pol99}.Amplitude and phase 
quadrature measurements are carried out respectively on the two output 
beams for each mode either through homodyne detection \cite{pol99} or 
parametric amplification \cite{ral99}. A classical channel for each of 
the polarization modes is formed from these measurements which are 
passed to the reconstruction site where they are used to displace the 
corresponding entangled pair for each mode. The output $c_{T}$ is 
formed by combining the two displaced polarization modes on a polarizing 
beamsplitter. Under conditions for which 
losses can be neglected the output from the teleporter is
\begin{equation}
c_{h,v,T}=\lambda c_{h,v}+(\lambda 
\sqrt{H}-\sqrt{H-1})f_{h,v,1}^{\dagger}+( \sqrt{H}-\lambda 
\sqrt{H-1})f_{h,v,2}
\label{tele}
\end{equation}
where $\lambda$ is the feedforward gain in the teleporter, the 
$f_{h,v,i}$ are vacuum inputs to the 2-mode squeezer providing the 
entanglement for the teleporter (see Fig.2(a)) 
and $H$ is the parametric gain of the 
squeezer. The fields are recombined in phase at the final 
beamsplitter giving the outputs
\begin{eqnarray}
a_{h,v,out} & = & {{1}\over{\sqrt{2}}}(c_{h,v,T}+d_{h,v})\nonumber\\
b_{h,v,out} & = & {{1}\over{\sqrt{2}}}(c_{h,v,T}-d_{h,v})
\end{eqnarray}
The photon counting rates of the two arms have expectation values
\begin{eqnarray}
<a_{out}^{\dagger}a_{out}> & = & 
\bra{\phi}_{a}\bra{\phi}_{b}({a_{h,out}}^{\dagger}+
{a_{v,out}}^{\dagger})(a_{h,out}+
a_{v,out})\ket{\phi}_{a}\ket{\phi}_{b}\nonumber\\
 & = & 0.25(1+\lambda)^{2}+(\lambda 
\sqrt{H}-\sqrt{H-1})^{2}\nonumber\\
<b_{out}^{\dagger}b_{out}> & = & 
\bra{\phi}_{a}\bra{\phi}_{b}({b_{h,out}}^{\dagger}+
{b_{v,out}}^{\dagger})(b_{h,out}+
b_{v,out})\ket{\phi}_{a}\ket{\phi}_{b}\nonumber\\
 & = & 0.25(1-\lambda)^{2}+(\lambda \sqrt{H}-\sqrt{H-1})^{2}
\label{exp}
\end{eqnarray}
In the limit of very strong entanglement squeezing 
($\sqrt{H}-\sqrt{H-1}\to 0$) we find from Eq. \ref{tele} that $c_{h,v,T}
\to c_{h,v}$ for unity gain ($\lambda=1$), 
i.e. perfect teleportation. For the same conditions (and only for 
these conditions) 
the visibility of the Mach-Zehnder outputs,
\begin{equation}
V={{<a_{out}^{\dagger}a_{out}>-<b_{out}^{\dagger}b_{out}>}\over{
<a_{out}^{\dagger}a_{out}>+<b_{out}^{\dagger}b_{out}>}}
\end{equation}
goes to one, indicating the state of the teleported arm exactly 
matches that of the unteleported arm. 
Notice that the expectation values (Eq.\ref{exp}), and 
thus the visibility, do not depend 
on the actual input state (no dependence on $x$ and $y$). Hence we 
can demonstrate that the teleporter is operating ideally even if we do 
not know the state of the input. Classical limits can be set by 
examining the visibility obtained with no entanglement ($H=1$). In 
Fig.3 we plot the visibility versus feedforward gain in the teleporter 
for the cases of no entanglement (0\%), 50\% entanglement squeezing and 90\% 
entanglement squeezing. Maximum visibility in the 
classical case is $0.42$. Increasing entanglement leads to increasing visibility.

It is known that using a single mode squeezed beam, divided in half on a 
beamsplitter (see Fig.2(b)), instead of a true 2-mode squeezed source 
(which exhibits 
Einstein, Podolsky, Rosen (EPR) correlations), can still 
produce fidelities of teleportation higher than the classical limit 
for coherent state inputs. 
Loock and Braunstein \cite{loc99} 
have recently contrasted various single mode and 
2-mode squeezing schemes on the basis of their fidelity. It is 
educational to examine how well the single squeezer teleporter performs in 
our single photon Mach-Zehnder. The input output relation for a single 
squeezer teleporter is
\begin{equation}
c_{h,v,S}=\lambda c_{h,v}+{{1}\over{\sqrt{2}}}((\lambda 
\sqrt{H}-\sqrt{H-1})f_{h,v,1}^{\dagger}+( \sqrt{H}-\lambda 
\sqrt{H-1})f_{h,v,1}+\lambda f_{h,v,2}^{\dagger}+
f_{h,v,2})
\label{tele2}
\end{equation}
The expectation values for the outputs then become
\begin{eqnarray}
<a_{out}^{\dagger}a_{out}> & = & 0.25(1+\lambda)^{2}+0.5(\lambda 
\sqrt{H}-\sqrt{H-1})^{2}+0.5 \lambda^{2} \nonumber\\
<b_{out}^{\dagger}b_{out}> & = & 0.25(1-\lambda)^{2}+
0.5(\lambda \sqrt{H}-\sqrt{H-1})^{2}+0.5 \lambda^{2}
\label{exp1}
\end{eqnarray}
On Fig.3 we also present the visibility as a function of gain for the 
single squeezer case with squeezing of 87.5\%. The squeezing is picked 
such that the average coherent state unity gain fidelity is the same 
as for the 50\% squeezed 2-mode entanglement (the criteria used in 
Ref.\cite{loc99}). The performance of the single squeezer teleporter is clearly 
inferior. Although achieving a better visibility than the 
classical teleporter it never exceeds, or equals, for any gain, the 
performance of the 50\% squeezed 2-mode teleporter. The maximum visibility of the 
2-mode teleporter is 25\% higher. We conclude that the entanglement of 
the single squeezer is not as useful for teleportation as might be 
suggested by the coherent state average fidelity measure.

In the experiments we have imagined so far the level of visibility 
has been determined not only by the ability of the teleporter to 
reproduce the input states of the photons (the mode overlap) but also 
the efficiency with which input photons to the teleporter 
lead to correct output photons (the 
power balance). It is of interest to try to separate these effects.
We can investigate just state reproduction if we allow 
attenuation to be applied to beam $d$, thus ``balancing" 
the Mach-Zehnder by compensating for the loss introduced by the 
teleporter (see Fig.1(b)). 
The attenuated beam $d$ becomes
\begin{equation}
d_{h,v,A}=\sqrt{\eta}d_{h,v}+\sqrt{1-\eta}g_{h,v}
\end{equation}
where $g$ is another vacuum field and $\eta$ is the intensity transmission of 
the attenuator. The expectation values of the outputs (using 2-mode 
entanglement) are now
\begin{eqnarray}
<a_{out}^{\dagger}a_{out}> & = & 0.25(\sqrt{\eta}+\lambda)^{2}+(\lambda 
\sqrt{H}-\sqrt{H-1})^{2}\nonumber\\
<b_{out}^{\dagger}b_{out}> & = & 0.25(\sqrt{\eta}-\lambda)^{2}+
(\lambda \sqrt{H}-\sqrt{H-1})^{2}
\label{exp2}
\end{eqnarray}
In Fig.4 we plot visibility versus gain, using the attenuation $\eta$ 
to optimize the visibility ($\eta \le 1$). Now we can always achieve unit 
visibility for any finite 
level of entanglement by operating at gain 
$\lambda_{opt}={{\sqrt{H-1}}\over{\sqrt{H}}}$ and balancing the 
interferometer by setting $\eta=\lambda_{opt}^{2}$. The high 
visibility is achieved because at gain $\lambda_{opt}$ the teleporter 
behaves like pure attenuation \cite{pol99}. That is the photon flux of the 
teleported field is reduced, but no ``spurious photons" are added to 
the field. Thus, at this gain, all output photons from the teleporter are in the 
right state, but various input photons are ``lost". This effect does 
not occur 
for the single squeezer teleporter (also plotted in Fig.4) 
whose performance is not improved 
by balancing the interferometer, further emphasizing its lack of 
useful entanglement. 

So far we have considered test arrangements in which a teleported 
field is compared with one which is not teleported. However the 
result of Eq.\ref{exp2} suggests a self testing arrangement for a 
teleporter. Suppose we place a teleporter in both arms of the 
interferometer as portrayed in Fig.1(c). Writing an expression for 
the teleported beams $d$ similar to Eq.\ref{tele} we find the 
expectation values of the outputs are now
\begin{eqnarray}
<a_{out}^{\dagger}a_{out}> & = & \lambda^{2}+2(\lambda 
\sqrt{H}-\sqrt{H-1})^{2}\nonumber\\
<b_{out}^{\dagger}b_{out}> & = & 
2(\lambda \sqrt{H}-\sqrt{H-1})^{2}
\label{exp3}
\end{eqnarray}
where we have assumed the gains of the two teleporters are the same. 
By monitoring the ``dark" output port ($b_{out}$) it may be 
possible to keep the system ``locked" to maximum visibility, without 
any knowledge of the input state or requiring the destruction of the 
output state ($a_{out}$). Once again, under low loss conditions, unit visibility 
is achieved for gain $\lambda_{opt}$ as illustrated in Fig.5. 
The added complexity of using two teleporters may be 
justified in practice by the greater versatility of this system.

In conclusion, we have examined a Mach-Zehnder arrangement for 
testing the efficacy of single photon qubit teleportation. The major 
advantage of this arrangement is it  
doesn't require the tester to know the input state of the photon. We 
have contrasted the results obtained with no entanglement, single 
mode entanglement and true 2-mode entanglement using continuous 
variable teleportation. The highest 
visibilities are always achieved with 2-mode entanglement. We have 
also suggested that a ``locked" teleporter could be constructed 
using a generalization of the testing scheme.  We have 
only examined here the case where losses can be neglected. Losses 
reduce visibilities but the general trends discussed here remain the 
same.

\begin{figure}
 \caption{Schematics of various Mach-Zehnder plus teleporter 
 arrangements.}
\end{figure}

\begin{figure}
 \caption{Schematic of the two types of entanglement used for 
 teleportation. NDOPO stands for non-degenerate optical parametric 
 oscillator and DOPO stands for degenerate optical parametric 
 oscillator. A separate pair of entangled beams is needed to teleport 
 each of the two polarization modes. Alternatively type II polarization 
 entanglement could be used [4]. }
\end{figure}

\begin{figure}
 \caption{Visibility versus gain for various levels of 2-mode 
 entanglement (0\%, 50\% and 90\%) and 87.5\% single mode squeezing 
 (single squeezer).}
\end{figure}

\begin{figure}
 \caption{Visibility versus gain with ``attenuation balancing'' 
 for various levels of 2-mode 
 entanglement (0\%, 50\% and 90\%) and 87.5\% single mode squeezing 
 (single squeezer). }
\end{figure}

\begin{figure}
 \caption{Visibility versus gain for self-testing teleporter 
 for various levels of 2-mode 
 entanglement (0\%, 50\% and 90\%). }
\end{figure}

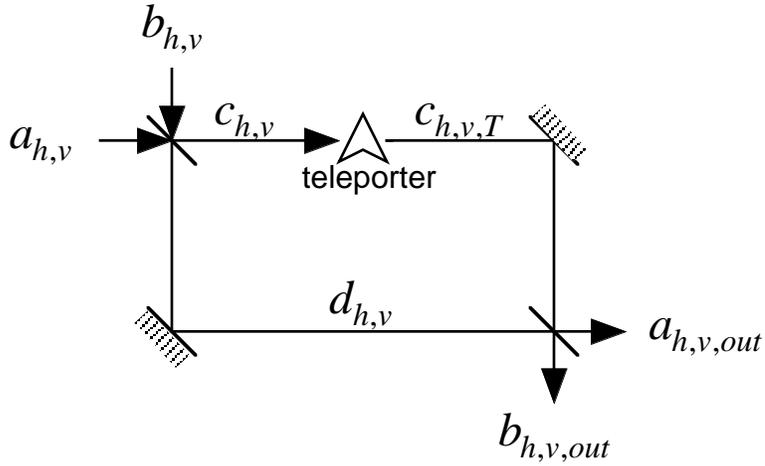
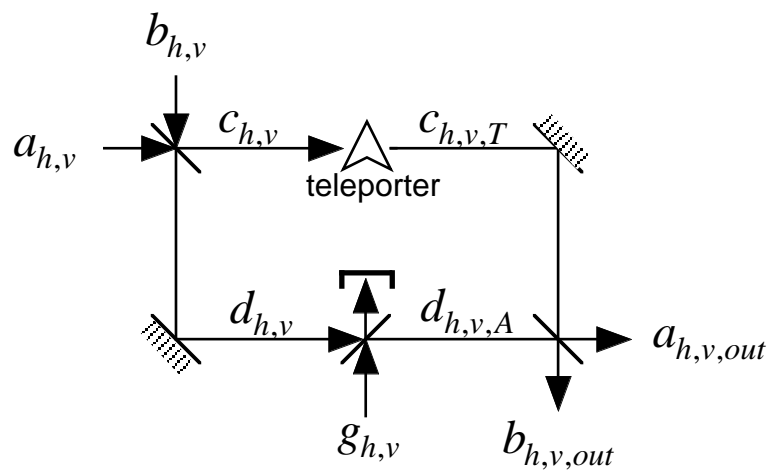
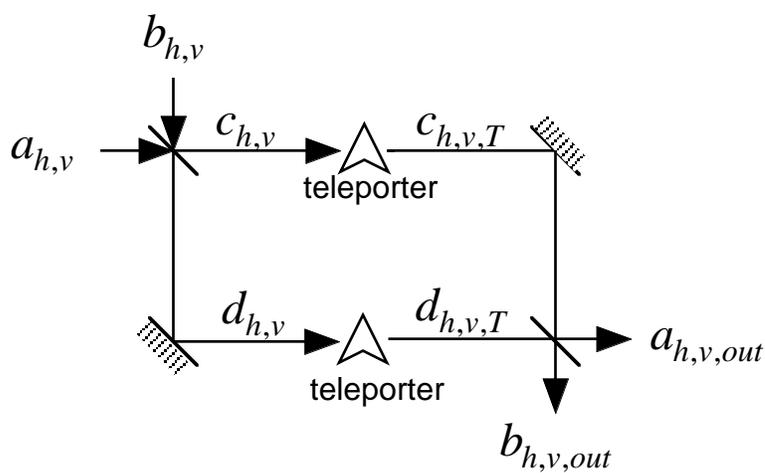
\end{document}